\documentclass[twocolumn,aps,floats,superscriptaddress,showpacs,floatfix,preprintnumbers]{revtex4}
\usepackage{epsfig}
\usepackage{graphicx}
\usepackage{amsmath}
\usepackage{amsfonts,amssymb}
\usepackage{color}
\usepackage{slashed}
\usepackage{ulem}   
\newcommand{\beq}{\begin{equation}} 
\newcommand{\eeq}{\end{equation}} 
\newcommand{\beqa}{\begin{eqnarray}} 
\newcommand{\eeqa}{\end{eqnarray}}

\begin{document}

%


\title{A condensed matter realization of the axial magnetic effect}
\author{Maxim N. Chernodub} 
\affiliation{CNRS, Laboratoire de Math\'ematiques et Physique Th\'eorique, Universit\'e Fran\c{c}ois-Rabelais Tours,\\ F\'ed\'eration Denis Poisson, Parc de Grandmont, 37200 Tours, France}
\affiliation{Department of Physics and Astronomy, University of Gent, \\ Krijgslaan 281, S9, B-9000 Gent, Belgium}
\author{Alberto Cortijo} 
\affiliation{Instituto de Ciencia de Materiales de Madrid, CSIC\\
Sor Juana In\'es de la Cruz 3, Cantoblanco, 28049 Madrid, Spain.}
\author{Adolfo G. Grushin}
\affiliation{Max-Planck-Institut $f\ddot{u}r$ Physik komplexer Systeme, 01187 Dresden, Germany}
\author{Karl Landsteiner} 
\affiliation{Instituto de F\'isica Te\'orica UAM/CSIC, \\
Nicol\'as Cabrera 13-15, Cantoblanco, 28049 Madrid, Spain }
\author{Mar\'{\i}a A. H. Vozmediano}
\affiliation{Instituto de Ciencia de Materiales de Madrid, CSIC\\
Sor Juana In\'es de la Cruz 3, Cantoblanco, 28049 Madrid, Spain.}

\date{\today}
\begin{abstract}
The axial magnetic effect, i.e., the generation of an energy current parallel to 
an axial magnetic field coupling with opposite signs to left- and right-handed fermions
is a non-dissipative transport phenomenon intimately 
related to the gravitational contribution to the axial anomaly. 
An axial magnetic field emerges naturally in  condensed matter in the so called 
Weyl semi-metals. We present a measurable  implementation of the axial magnetic effect.
We show that the edge states of a Weyl semimetal 
at finite temperature possess a temperature dependent angular momentum in the direction 
of the vector potential intrinsic to the system. Such a  realization provides a plausible
context for the experimental confirmation of the elusive gravitational anomaly.

\end{abstract}
%
\preprint{IFT-UAM/CSIC-13-120}
%
%
%
 \maketitle


Anomalies have played an important role in the construction of consistent quantum
field theory (QFT) and string
theory models. Among them, the most intensively investigated case 
is that of the axial anomaly, responsible
for the decay of a neutral pion into two photons \cite{Ioffe06}. Similarly 
in a curved space gravitational anomalies \cite{AW84} can occur and
mixed axial-gravitational anomalies give rise to very interesting predictions as the decay
of the  pion into two gravitons. While the former phenomenon is by now well established there are
so far no  experimental settings to provide evidence of the gravitational anomaly.
More recently anomalies are starting to play
an interesting role as responsible for exotic transport phenomena in QFT in extreme conditions.
In the context of the quark gluon plasma \cite{Satz11} it has become clear
in recent years that  at finite temperature and density quantum anomalies
give rise to new non-dissipative transport phenomena.

The recognition of the role
of topology in the classification of condensed matter systems started long time ago with the prototypical example set by liquid helium \cite{Vo03}. The low energy excitations of $He_3$ are described by Dirac fermions what made the system an interesting analog to study  high energy phenomena. The advent of 
new materials (graphene, topological insulators and superconductors \cite{HK10,QZ11}, Weyl semimetals) whose low energy electronic properties are described by Dirac fermions in one, two or three spacial dimensions  
has enlarged and widened the analogy between high energy and condensed matter in this century.  
Simultaneously, the new experiments on the quark--gluon plasma
and the recent developments in holography have opened an unexpected scenario where
high energy and condensed matter physics merge. In this work we  propose a condensed matter scenario for the experimental realization of the gravitational anomaly.

The most commonly cited example of the new non-dissipative transport phenomena occurring in the quark--gluon plasma
is the chiral magnetic effect \cite{FKW08} that refers to the generation of an electric
current parallel to a magnetic  field whenever an imbalance between the number of right 
and left-handed fermions is present. Another interesting example
is the axial magnetic effect (AME) associated with the generation of an energy current parallel to
an axial magnetic field, i. e. a magnetic field coupling with opposite signs to right and left handed
fermions. As will be described later, the AME conductivity has a contribution proportional to the temperature which is directly related to the gravitational anomaly. 
In this Letter we argue that although an axial gauge field is absent in nature at a microscopic/fundamental level,
it can easily appear in an effective low energy theory
describing Weyl semi-metals. We also provide a
possible setup to ascertain the temperature dependent
component of the AME conductivity, which directly probes the gravitational anomaly.

{\it Weyl Semi-metals.}~
To set up the notation used throughout the paper and for completeness
we will here explain the  low energy description
of Weyl semi-metals \cite{XTVS11,BB11}.
As a working definition, 
Weyl semi-metals are materials for which their
 low energy degrees of freedom are 
described by  (3+1)D Weyl fermions, i.e.  two component spinor 
solutions of the Weyl Hamiltonian
%
\label{eq:Weyleq}
$H_{\mathbf{k}}=\pm\boldsymbol{\sigma}\cdot\mathbf{k}$.
%
In 3+1 dimensions, $\boldsymbol{\sigma}=(\sigma_{x},\sigma_{y},\sigma_{z})$
and $\mathbf{k}$ is the three-dimensional momentum. 
The $\pm$ signs 
correspond to the two 
chiralities (left and right) of a Weyl spinor. In lattice systems Weyl fermions must appear in 
pairs of opposite chiralities due to the 
Nielsen-Ninomiya  theorem \cite{NielNino81a}. They will be in general separated 
in momentum space and also shifted in energies. Since they  can only annihilate 
in pairs, the separation in momentum space endows 
the nodes with a notion of topological stability. \cite{XTVS11,BB11} \\
Experimentally, magnetically doped Bi$_2$Se$_3$ and TlBiSe$_2$ \cite{CCA10,WXX11,CZL13} can be a feasible route
to realize the Weyl semi-metal phase \cite{XPQ13}. 
Several band structure calculations
\cite{YTK12,WSY12,WWW13,SYZ13}
have predicted band touching to occur in Cd$_3$As$_2$ and A$_3$Bi (A=Na, K, Rb) compounds.
Very recently there has been remarkable experimental evidence \cite{BGetal13,Netal13,LZW13} of such
3D Dirac semimetal state in this family of materials, which under magnetic doping would potentially
host the Weyl semi-metal phase.  In addition, heterostructures that alternate between trivial and topological insulators
have been realized experimentally \cite{KET12}. These could be coated with ferromagnetic insulators as was experimentally demonstrated for a single layer \cite{WKA13}, closer in spirit to the early proposal considered in Ref. \onlinecite{BB11}. 
This latter case realizes the minimal number of nodes (two) and so 
the characteristics of the Weyl semi-metal phase
 are taken into account 
by the following low energy action \cite{KV05,ZWB12,Gr12,ZB12,GT12}
\begin{equation}
\label{eq:Weylact}
S=\int d^4k \bar{\psi}_{k}(\gamma^{\mu}k_{\mu}-b_{\mu}\gamma^{\mu}\gamma_{5})\psi_{k},
\end{equation}
where $k^{\mu}=(k_{0},\mathbf{k})$ is the momentum four-momentum, $b_{\mu}$ 
is a constant four-vector and $\psi_{k}$ is a four component spinor. The vector $b_{\mu}$ has physical origin; the spatial part $\mathbf{b}$ breaks time-reversal symmetry and preserves inversion and can be induced by 
doping the system with magnetic impurities \cite{BB11}. The time-like component $b_{0}$ on the other hand 
breaks inversion and preserves time reversal and can be originated in a particular spin-orbit coupling term. \cite{ZB12} 
As a consequence, 
the energy spectrum for this case results in two Weyl nodes separated by
$\Delta\mathbf{k}=2\mathbf{b}$ and $\Delta E=2b_{0}$ in momentum 
and energy respectively ($\Delta k^{\mu}=2b^{\mu}$ in a more compact notation). 
In this work we will take this action as a starting point to describe
the low energy physics of Weyl semi-metals. \\
\begin{figure}
\includegraphics[scale=0.25,page=1]{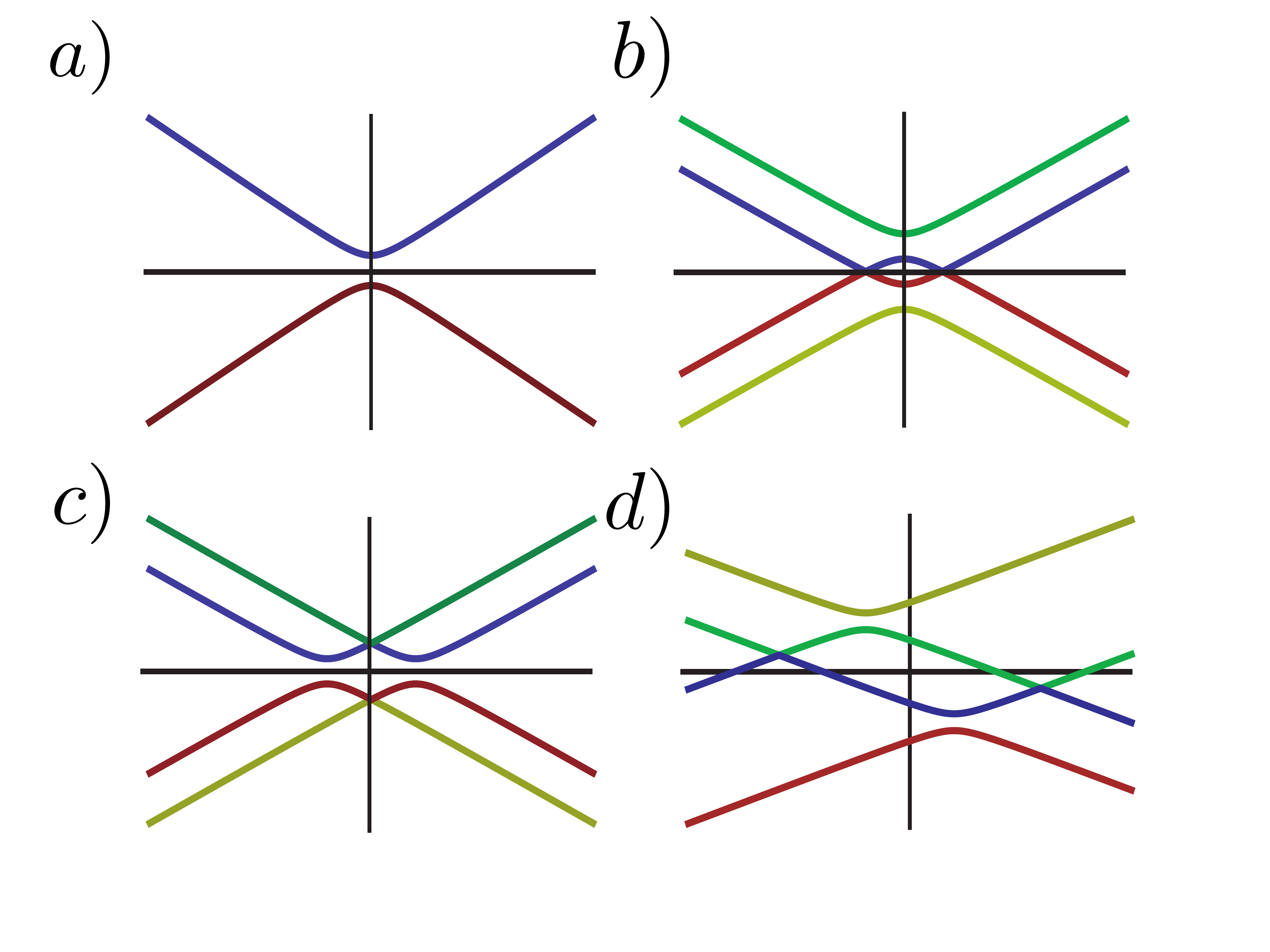}
\caption{\label{Fig:WSMbands} (Color online) Dispersion relation from \eqref{eq:Weylactm} with $m\neq 0$ 
and (a)$-b^2<m^2$, including   $b^{\mu}=0$, (b) $b^{i}\neq 0$ and $b_{0}=0$ with $-b^2>m^2$, (c) $ m\neq 0,b^{i} =0$ 
and $b_{0}\neq 0$. This case always satisfies $-b^2<m^2$ for any value of $b_{0}$. (d) $b^{\mu}\neq 0$  with $-b^2>m^2$. }
\end{figure}  
Before doing so, it is important to address the regime of validity for this action.
In real Weyl semi-metal materials, the two chiralities of the Weyl fermions 
will mix at higher energies above
some characteristic energy scale $m^2$, when $b^2\sim m^2$ with $b^2=b^2_{0}-\mathbf{b}^2$. Above this energy, 
a different band structure takes over and one cannot model the system with isolated Weyl nodes.
In that case a simple extension of 
\eqref{eq:Weylact} can be used to take into account the new energy scale, in particular \cite{Gr12} 
\begin{equation}
\label{eq:Weylactm}
S=\int d^4k \bar{\psi}_{k}(\gamma^{\mu}k_{\mu}-m-b_{\mu}\gamma^{\mu}\gamma_{5})\psi_{k}.
\end{equation}
which can be directly derived from microscopic models \cite{Gr12,VF13} and resembles a Lorentz breaking QED action \cite{Gr12}. 
Contrary to naive expectation the spectrum of \eqref{eq:Weylactm} need not be gapped even when $m\neq0$, 
i.e. the attributes gapless and massless are no longer interchangeable. 
In fact, whenever the condition $-b^2 < m^2$ is satisfied, the spectrum is gapped and the system is an insulator \footnote{Note the important point that when $b^{\mu}$ is purely timelike, the spectrum is gapped for any value of $b_{0}$. This rules out the possibility of the chiral magnetic effect for a purely timelike $b^{\mu}$. \cite{Gr12}}.
In the opposite case when $-b^2 > m^2$, the spectrum is gapless and contains two nodes 
separated both in momentum and energy (see Fig.~\ref{Fig:WSMbands}). Therefore in the latter case
 the material realizes the Weyl semi-metal phase.
The separation between nodes in the latter case is proportional to the four-vector 
$\Delta k^{\mu} \sim b^{\mu}\sqrt{1-\frac{m^2}{b^2}}$.\\
Both \eqref{eq:Weylact} and \eqref{eq:Weylactm} lead to interesting predictions
such as the presence of surface states in the form of Fermi arcs \cite{XTVS11,BB11}, a Hall response \cite{BB11}
as well as a current response parallel to an external 
magnetic field \cite{ZWB12,Gr12,ZB12,GT12}, an analogue of the the chiral 
magnetic effect \cite{FKW08}, although the realization of the latter is still under active 
debate \cite{Gr12,ZB12,GT12,VF13,CWB13,BKY13,Land13}.

The exact magnitude of $m$
will depend on the particular realization of the Weyl semi-metal phase.  A way to estimate 
its magnitude is by realizing the Weyl semi-metal phase by closing the gap of a topological insulator simply by adding magnetic 
impurities which break time-reversal symmetry \cite{XPQ13}. 
In the process, the single particle gap of the topological 
insulator $m$ closes while interpolating from a situation with $-b^2 < m^2$ to the Weyl semi-metal phase with 
$-b^2 > m^2$. In this simple picture the value of $m$ is as large as the gap of the original topological insulator 
which for Bi$_{2}$Se$_3$ is close to $\sim 0.3$eV \cite{Zhang09,X09,HMHasan09}. This gives an upper bound for $m$ 
although in general one can expect it to be smaller. \\

We now address the question of how the low energy description \eqref{eq:Weylact} generates a finite AME in a  Weyl semimetal. 
The axial magnetic effect describes the generation of an {\it energy} current parallel to an axial magnetic field ${\bf B}_5$ (i. e. a magnetic field coupling with opposite signs to left and right fermions) in a system of massless Dirac fermions in 3+1 dimensions at finite temperature and chemical potential
\beq
T^{0i}=J^i_\epsilon=\sigma_{AME} B^i_5.
\label{ame}
\eeq
\begin{figure}
\includegraphics[scale=0.25,page=2]{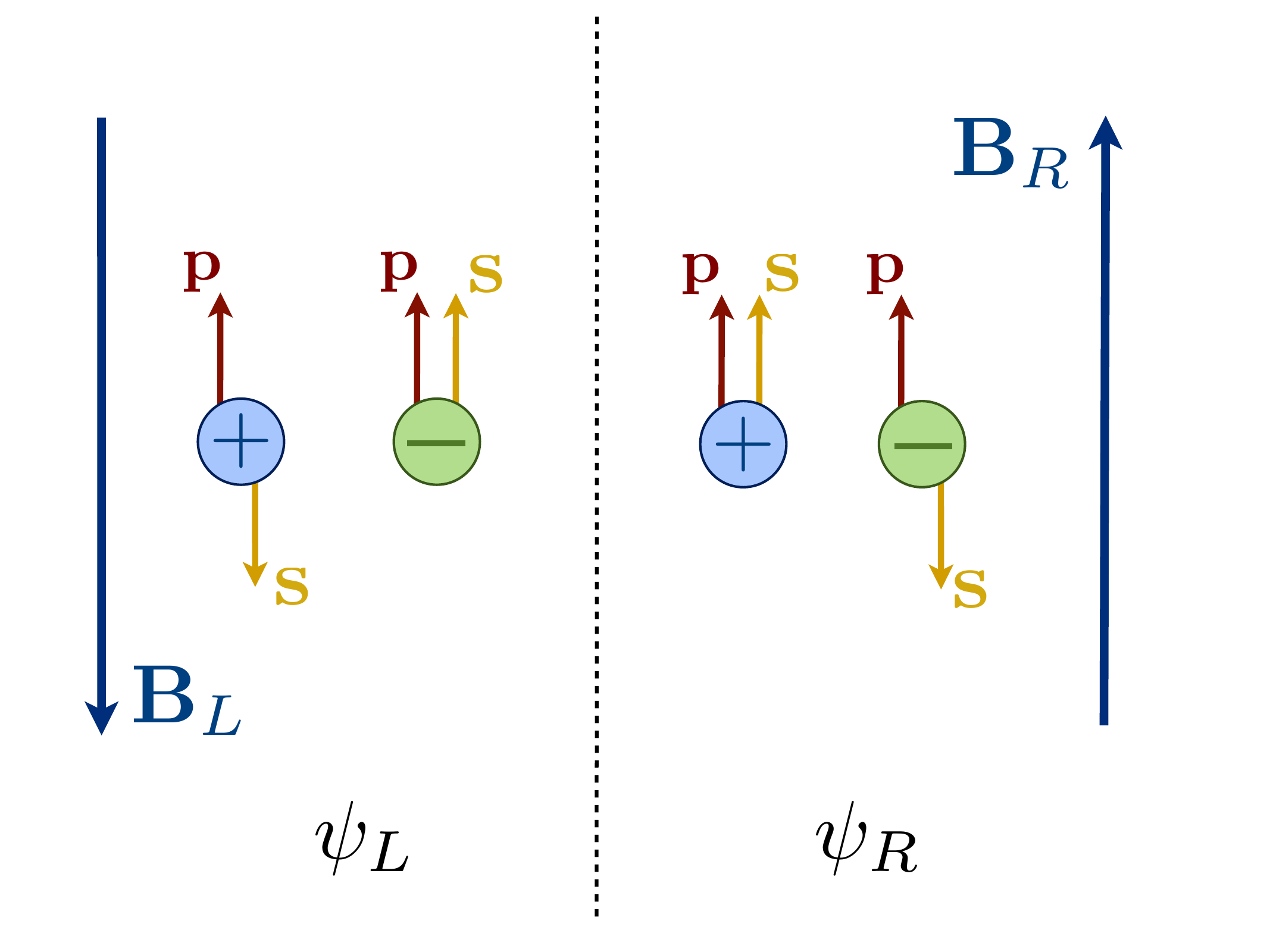}
\caption{\label{LLL} (Color online) The lowest Landau level picture of the axial magnetic effect.
Here, L and R represent the two different chiralities, $\pm$ indicate the charge of the particle/antiparticle with momentum $\mathbf{p}$. The vector $\mathbf{s}$ shows the direction of spin for each type of particle.
 }
\end{figure}  
A Landau level picture of this effect can be obtained by adapting
the derivation of the chiral magnetic effect in terms of Landau levels
 done in
Ref. \onlinecite{AKW10}. Fig. \ref{LLL} shows a schematic view of the effect.  The spectrum
of massless Dirac fermions in an axial magnetic field is organised into Landau levels. In
the lowest Landau level the spins and momenta are aligned according to chirality. The
chiral magnetic field acts with a relative sign on right- and left-handed fields $\Psi_{R;L}$. The
particle quanta of the right-handed field have their spin aligned with the magnetic field,
whereas the anti-particle quanta of $\Psi_R$ have their spin anti-aligned. 
For the quanta of the left-handed 
field these relations are reversed as is the sign of the magnetic field. Therefore in the absence of
any imbalance of either charge or chirality all quanta have their momenta aligned in the
background of an axial magnetic field and create an energy flux in the direction of the chiral magnetic
field. The  energy flow is higher the more quanta are on shell, i.e. the
higher the temperature. The
higher Landau levels are degenerate in spin and therefore their overall momenta average
out to zero. 

From Refs. \onlinecite{LMP11,Vi80} it follows that for a single (massless) Dirac fermion, i.e. one pair of
Weyl-cones, the axial magnetic conductivity  is
\beq
\sigma_{AME}=\frac{\mu^2+\mu^2_5}{4\pi^2}+\frac{T^2}{12},
\label{sigmaAME}
\eeq
where $T,\mu$ and $\mu_{5}$ are the temperature, chemical potential and axial chemical potential respectively. 
Of particular interest is the fact that this conductivity has a purely temperature
dependent contribution, i.e. even at zero density and in the absence of a chiral imbalance, the
AME is not zero. The coefficient of the $T^2$ term can  be inferred from purely hydrodynamic arguments \cite{SS09,NO11,JLY13}, and has  been computed recently in lattice simulations of
QCD \cite{BCetal13,Bui13}. This temperature dependence is a direct 
consequence of the presence of the (mixed) axial--gravitational anomaly \cite{LMP11,JLY13}:
\begin{equation}
\label{eq:gravanomaly}
\partial_\mu J^\mu_5 = \frac{1}{384\pi^2} \epsilon^{\mu\nu\rho\lambda} 
R^\alpha\,_{\beta\mu\nu} R^\beta\,_{\alpha\rho\lambda}\,,
\end{equation}
where $J^\mu_5$ is the axial current and $R^\alpha\,_{\beta\mu\nu}$ is curvature tensor.

Next we show the generation of a net angular momentum carried by the surface states (the Fermi arcs) of a Weyl semi-metal due to the AME. 
The simplest action capturing the features of a neutral ($\mu=\mu_{5}=0$) Weyl semi-metal is \eqref{eq:Weylact}
\begin{equation}
S=\int d^4k \bar{\psi}_{k}(\gamma^{\mu}k_{\mu}-b_{\mu}\gamma^{\mu}\gamma_{5})\psi_{k},
\end{equation}
where $b_{\mu}$ acts as a chiral gauge potential. For a Weyl semi-metal $b_{i}$ is constant in the bulk and goes to zero sharply at the edge
so there will be a strong effective axial magnetic field ${\bf B}_{5}=\nabla\times {\bf b}$ induced there.
This in turn implies that through the AME, $T^{0i}$ can generate a finite angular momentum for the states at the boundary.
Consider a cylinder of Weyl semimetal of height $L$ and basal radius $a$ with the simplest configuration 
$\mathbf{b}=\hat{z}b_{z}\Theta(a-|r|)$. The axial magnetic field will point in the azimuthal direction and be proportional to $B_\theta\sim b_z\delta(|r|-a)$. The corresponding energy current \eqref{ame} will induce an angular momentum $L_k=\int_{\cal V} \varepsilon_{ijk} x_i T_{0j}$ along the axis of the cylinder (of volume ${\cal V}$):
\beq
\label{lz}
L_{z} =\int_{\cal V} \varepsilon_{z r \theta} \;r\;T_{0\theta}= 2\pi \sigma_{AME} a^2\;L b_{z}.
\eeq

Plugging in the expression  \eqref{sigmaAME} for $\sigma_{AME}$ it follows that, at zero density and in the absence of a chiral imbalance the states at the edge of the cylinder posses an angular momentum of magnitude
\beq
\label{Lz}
L_z=\frac{N_f}{6} T^2 b_z {\cal V},
\eeq
where $N_f$ is the number of pairs of Weyl cones, and $b_z$ is the effective axial potential proportional to the separation of the Weyl cones in momentum space \cite{BB11}.

Notice that, although we have assumed a constant $b_{z}$  to simplify the formulas, 
any spacial variation of 
$b_{i}$ in the bulk would give rise to an effective axial magnetic field and to an energy current supported in the bulk. This is an important difference with previous models \cite{Kit06,KV05} where the current is intrinsically an edge current. In a physical realisation of the type discussed in \cite{BB11}, the axial field originates in the 
magnetization of the induced dopants and can be easily chosen to be inhomogeneous.

A direct observation of  the rotation is hindered by the fact that only the edge states carry  the angular momentum and the dissipationless rotation will not drag with it the ions of the lattice. As explained below, the distinctive characteristic that might allow its detection is the explicit  $T^{2}$ coefficient coming from the axial magnetic effect.

\begin{figure}
\includegraphics[scale=0.3]{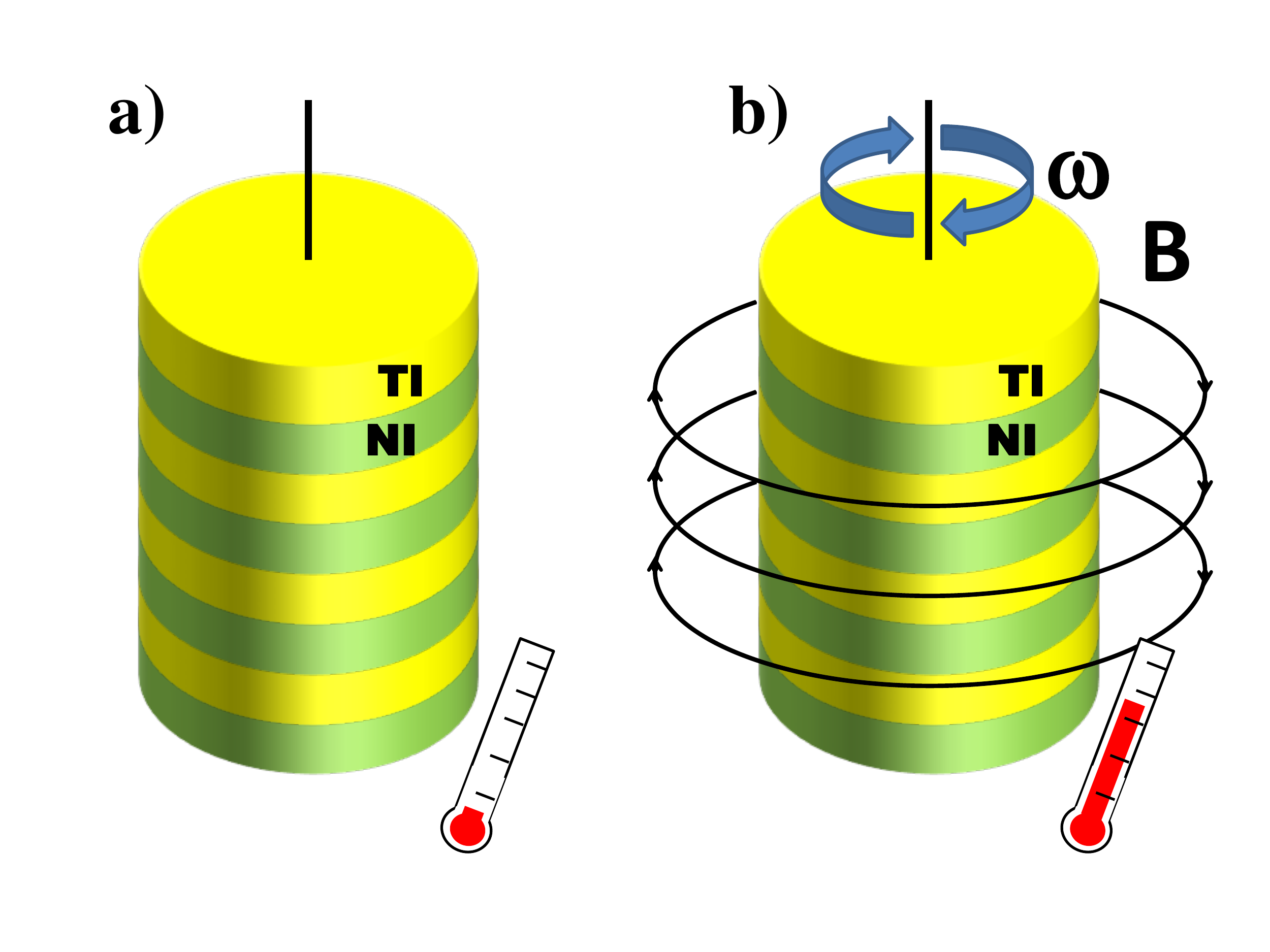}
\caption{\label{rotation} (Color online) Sketch of the proposed setup to measure the rotation described in the text. A Weyl semi-metal will rotate under heating or cooling through the axial magnetic effect.} 
\end{figure}

To be specific we will focus on the physical realization of a Weyl semi-metal proposed in Refs. \onlinecite{BB11,ZWB12} discussed above. It has two Weyl modes, although our proposal is extensible to other possible realizations of this phase with a larger number of Weyl nodes.
In Ref. \onlinecite{Gr12} it was shown that the low energy action of the model in real space is the action \eqref{eq:Weylactm} which reduces to \eqref{eq:Weylact}.\\
Consider the cylinder  in isolation and suspended as sketched in Fig. \ref{rotation}.  

If the system initially at a given temperature $T_{i}$, is heated to $T_{f}=T_{i}+\Delta T$ the angular momentum due to the AME will increase. 
Since the total angular momentum is conserved, the cylinder has to rotate in the opposite direction to compensate. 
The change in angular velocity is given by the change in angular momentum $\Delta L _{z}= I\Delta \omega_{z} $ through
\begin{equation}
\Delta \omega_{z}= - \frac{N_f b_z}{3\rho a^2} (2 T \Delta T + \Delta T^2),
\end{equation}
where we used that the moment of inertia of the cylinder of mass $M$ is $I=\frac{1}{2}Ma^2$.

The magnitude of ${\bf b}$ is determined by the expectation value of the magnetization of the induced dopants and can be estimated to be of order $0.01-0.1$ eV. Restoring the appropriate constants $\hbar$ and $v_F\sim 10^{-3}c$ and for  conservative values of  the magnitudes $a=1$mm, T=10K, $\rho$=10 g/cc, we get an estimate of the angular velocity of $\omega\sim 10^{-13}-10^{-11} s^{-1}$. Although small, this rotation can in principle be detected by standard optical devices \cite{SS88} or torque experiments. The angular velocity can increase  considerably by increasing the temperature interval but this is bound by the magnetic structure of the Weyl semimetal. Insulating ferromagnets such as the rare earth oxide EuO have Curie temperatures of 60-70 K however, a recent a recent publication \cite{OC12} showed that antiferro or ferrimagnetic materials can also be used to obtain the Weyl semimetal what would allow to increase the values up to room temperature. It is also to be noticed that the Fermi velocity will play the role of the speed of light in the conversion factors. A lower value of $v_F$ greatly enlarges the angular velocity . This is expected since the present effect is of thermal origin. Thus, for small values of $v_F$  it is easier to thermally populate states with higher momentum $p$.
Since $p$ determines the energy current and thus the angular momentum
it is reasonable to expect that the effect gets enhanced as $v_F$ becomes smaller
simply because it costs less energy to populate states with higher $p$.

The spontaneous generation of angular momentum and an edge current are typical phenomena in parity-violating physics as occurs for instance in the A phase of helium-3 \cite{Vo03} (see also \cite{Sauls11}). Our model adds a great versatility to these cases since in the Weyl semimetal case one can construct a lattice model with a 
spatially dependent $b_{i}$ vector in real space, without
any need to invoke distance between Fermi-points \cite{BB11}. The space variation of the axial field is linked to the distribution of the magnetic impurityes and can be easily manipulated to design an experiment.

A similar energy current  with 
temperature scaling $T^2$ as the one described in this work was obtained in \cite{Kit06} in a two dimensional model but this
is intrinsically 
formulated as an edge current while ours is a bulk effect. Although we have chosen for simplicity
an example where the effective axial magnetic field only exists at the edge of the sample, in our case
any spacial variation of $b_i$ will give rise to an energy current with support in the region
where the axial magnetic field is non zero. On a more formal level it is worth noticing that  the existence of an energy current in \cite{Kit06} is traced back to the presence of a 2-dimensional pure gravitational anomaly whereas in our work the current is due to the 4-dimensional mixed gauge-gravitational anomaly, which is the deeper reason why the axial magnetic effect is essentially a bulk phenomenon.

The anomaly related responses discussed in this work are very hard to measure in the context of the quark-gluon plasma. Although there are indirect indications of the observation of the chiral magnetic effect, for instance in the ALICE detector of the LHC \cite{ALICE13}, at the moment there are no proposals for experiments that can directly observe the axial magnetic effect \cite{KS10}. This is due to the absence of axial magnetic fields in the high energy experiments. In this sense it is interesting to note that these are quite common in the effective low energy models of condensed matter systems.  An axial magnetic field arises from lattice deformations in graphene in (2+1) dimensions from which a mixed gravitational--deformation anomaly effect has been  proposed   recently \cite{VAetal13}.

A nonzero angular momentum density has been described recently in a three dimensional conformal field theory
\cite{LOSY13} within a holographic model. 
A dimensional reduction of the system proposed here 
will probably give the same result  providing a backup for the
somewhat obscure holographic ideas. 

To conclude we have shown that the AME gives rise to rotation in  a Weyl semi-metals upon heating or cooling due to an intrinsic axial magnetic field present in these systems.  This effect is determined by the thermal component of the AME, a direct consequence of the elusive gravitational anomaly, impossible to probe in high energy contexts.

We thank F. Guinea for discussions and Jens H. Bardarson for the critical reading of the manuscript.
Special thanks are given to G. Volovik for calling our attention to previous works on the subject.
This research was supported  by following grants:  FIS2011-23713, PIB2010BZ-00512,
FPA2012-32828, CPAN (CSD2007-00042); HEP-HACOS S2009/ESP-1473, SEV-2012-0249.  
ANR-10-JCJC- 0408 HYPERMAG. A. C.  acknowledges the JAE-doc European-Spanish program.


\newcommand{\npb}{Nucl. Phys. B}\newcommand{\adv}{Adv.
  Phys.}\newcommand{\epl}{Europhys. Lett.}

\end{document}